# Submandibular Sialolithiasis in a 9-year-old child: case report

Benhoummad Othmane[1], Rami Mohammed[2]*, Zaoual Atmane[2], Youssef Rochdi[2], Abdelaziz Raji[2]

1 Faculty of Medicine and Pharmacy of Agadir, Ibn Zohr University, Agadir, Morocco

2 Faculty of Medicine and Pharmacy of Marrakech, Cadi Ayyad University, Marrakech, Morocco

**Abstract**

*Sialolithiasis rarely occurs in children; it is observed more commonly in adults. Various treatment modalities for sialolithiasis have been reported in literature; we report the case of 9 years old child, with no particular pathological history. For 4 years, he had a tumefaction of the left submandibular region, associated with painful blocking episodes of salivation and an increase in the size of the tumefaction (accentuated during meals). clinical examination found a slight painless tumefaction of the left sub-mandibular region, with no pus or blood outflow, no endobuccal expression and no palpable calculi ultrasound examination shows a heterogeneous left submandibular gland, discreetly vascularized by Doppler, with multiple calculi. the patient was operated by cervical approach, a submandibularectomy was performed after detachment of the musculocutaneous flaps, preservation of the artery and the facial vein and the lingual nerve, and dissection of the submandibular canal. Anatomo-pathological study was performed and it showed macroscopically a yellowish enclosed calculus of 3mm of great axis. Microscopically, it shows a salivary glandular tissue with no signs of malignity, in this article, clinical findings, etiologies and different treatment approaches were reviewed .*

**Keywords:** Salivary glands; Submandibular; Calculi; Sialadenitis; Children

**Introduction:**

Sialoliths are calcareous concretions that may be found in the ducts of the major or minor salivary glands or within the glands themselves. They are thought to form by deposition of calcium salts around a central nidus which may consist of desquamated epithelial cells, bacteria, products of bacterial decomposition, or foreign bodies (1). The disease entity is known as sialolithiasis and is a rare occurrence in children. about 3% of all sialolithiasis cases are in children The condition is found more commonly in middle-aged adults. And The salivary gland most commonly affected is the submandibular gland (2)

**Case presentation:**

The patient was 9 years old, with no particular pathological history. For 4 years, he had a swelling of the left submandibular region, associated with painful blocking episodes of salivation and an increase in the size of the swelling (accentuated during meals). without fistula or inflammatory sign or reflex otalgia and without any other ENT or extra-ENT sign, all evolving in a context of apyrexia and conservation of the general state. clinical examination found a slight painless tumefaction of the left sub-mandibular region, with no pus or blood outflow, no endobuccal expression and no palpable calculi, not associated with palpable adenopathy.

ultrasound examination shows a heterogeneous left submandibular gland, discreetly vascularized by Doppler, with multiple calculi generating a posterior shadow cone, the largest of which measures 4.4 x 5mm (fig1).

Sialendoscopy was not available so we opted for surgery, the patient was operated by cervical approach at 2 fingerbreadths under the mandibular angle, a submandibularectomy was performed after detachment of the musculocutaneous flaps, with preservation of the mental branch of the facial nerve, the artery, the facial vein and the lingual nerve, and dissection of the submandibular canal.

The immediate and short-term postoperative follow-up was simple. Anatomo-pathological study was performed and it showed macroscopically a fragment of 6 g measuring 3x3x1cm (fig3), it presents to the section a yellowish enclosed calculus of 3mm of great axis (fig 4). Microscopically, it shows a salivary glandular tissue with no signs of malignity. The glandular parenchyma is dissociated by fibrous tracts and shows an abundant lymphoplasmacytic chronic inflammatory infiltrate in the absence of granulomatous signs or malignancy.

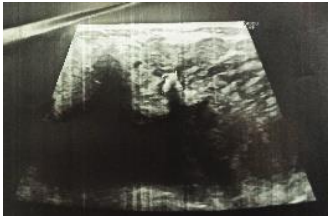

Fig1: Ultrasonography showing the main calculus.

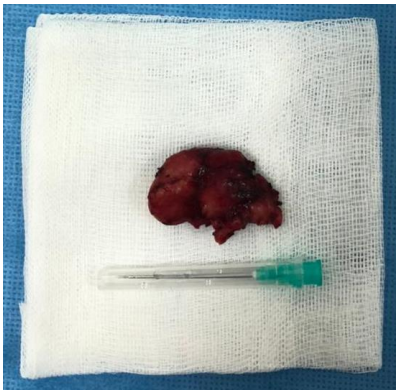

Fig3: the excision piece

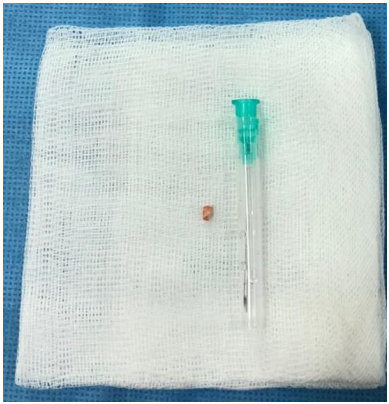

Fig4: the main calculus

**Discussion:**

Sialolithiasis is a relatively common condition in adults but is rarely found in children(2). it is revealed most often by a swelling of the submandibular region, associated with painful blocking episodes of salivation, slight tenderness in the floor of the mouth, and rarely pus outflow (3)(4).

When occurring, submandibular calculi are frequently formed in submandibular duct. The commonest sites are where the duct turns round the distal edge of the mylohyoid muscle, as the duct crosses the lingual nerve and just distal to the duct orifice(5). In our patient the calculi were situated inside the submandibular gland.

The patient's history, together with symptoms and careful clinical examination in addition to ultrasound examination may be sufficient to support the diagnosis as in our case. Orthogonal and distal oblique occlusal radiographs are especially useful(6). An orthogonal occlusal radiograph can show a calculus in the anterior portion of the Wharton's duct, while a stone in the posterior portion of the duct, a stone close to the hilum or in the gland can be demonstrated by a distal oblique occlusal radiograph (7). In general, 20–40% of salivary stones are radiolucent (8). Other investigations including sialography, CT scan and MRI may be necessary to localize a stone (9) (10) (11) (12)

Extraoral transcutaneous ultrasonography has been shown to be a safe and simple technique for the detection of glandular calculi. It has the advantage of avoiding the use of ionizing radiation and being able to detect radiolucentcalculi. However, the ductal area may be difficult to visualize by this approach(13) (14).

Sialography may be used to diagnose a stone, but has some disadvantages due to cannulation of a duct and retrograde injection of contrast material. If a calculus is present at the duct orifice, the cannulation may be impossible or contraindicated because of the risk of calculus displacement. When a gland infection is already established, X-ray sialography is contraindicated because of risk of extravasation. Intraoral ultrasound is not widely practiced. This is why in some cases when obstructive disorders or calculi are suspected without confirmation by X-ray or extraoral ultrasound, CT scan or MRI can help to confirm a diagnosis and localize a stone. To avoid radiation in children, MRI is theoretically preferable to CT scan, although sedation may be necessary for some of these examinations in children. During the acute phase of clinical obstruction, therapy is supportive and includes antibiotics, analgesics, and forced hydration. Hyperhydration can also help spontaneous elimination of small calculi. Once the acute stage subsides, definitive treatment can be instituted. Generally, symptomatic stones are removed surgically. However, non-invasive techniques used in selected cases have included shock-wave lithotripsy and endoscopic laser lithotripsy (15) (11). Occasionally, spontaneous exfoliation of a stone through the ductal orifice may occur(16) . If these treatments fail, especially if a calculus is located in the gland itself, extirpation of the affected gland is the treatment of choice(7)(5), and is also indicated in patients with recurrent sialolithiasis (7) . Many authors agree with the surgical removal of sialoliths situated anteriorly (17) (18) (19). Nevertheless, it is not the same when these are situated posteriorly or if the stones are recurrent. When it has been established that the calculi are intraglandular, it is generally recommended that the entire gland with the stone be removed (9) (20)(8). Until now, there is no known effective preventive treatment, but we recommend hyperhydration of all patients after surgical treatment.(17)

Sialendoscope is a good diagnostic tool for ductal pathology and unlike other radiological procedures findings of sialendoscopy correlate fairly with the symptomatology. it can be used both for diagnostic as well as therapeutic purposes and very often it can be done in a single sitting. Though it is also an invasive procedure, morbidity associated with sialendoscopy is mostly minor and that too most of the time is temporary.

Currently, sialendoscope is being most commonly used for removing submandibular and parotid duct calculi which is the most common obstructive pathology affecting the salivary glands

Smaller stones can be removed through the sialendoscope itself. Larger stones can either be broken down with extracorporeal shockwave lithotripsy (ESWL) or with laser fiber passed through the scope (intracorporeal lithotripsy). Alternatively, these larger stones can be removed with combined approach (sialendoscope guided external approach)

Juvenile recurrent parotitis (JRP) is recurrent inflammatory condition of the parotid glands affecting the paediatric population, with unknown aetiology. Sialendoscopy in these patients helps in understanding the pathology, ruling out any localized cause and it has a therapeutic role also. Steroids which are being used systemically for this pathology can now be delivered directly to the desired site, through the scope. This increases the efficacy and reduces the systemic side effects of the steroid treatment, reflecting in overall improved control and/or cure rate.

Ductal strictures which may be secondary to ductal calculus can be very effectively treated with the help of sialendoscopes. Sialendoscopy helps in direct and precise evaluation of the nature, site and length of the ductal strictures and consequently the optimal intervention.

Sialendoscopy should not be attempted during the acute inflammation of salivary glands. It may increase the pain and swelling in an already inflamed gland. Inflammation results in decreased cannulation rate, poor visibility of the ductal system which may result in complication like perforation of the duct leading to stenosis of the duct, thus increasing the overall failure and complication rate.(21)

**Conclusion:**

Sialolithiasis is not commonly observed in children, but should be considered in the differential diagnosis in patients with submandibular swelling and pain. Establishing a diagnosis of sialolithiasis requires a thorough history and physical examination along with paraclinical evaluation, ultrasound is the safest and simplest tool and could be sufficient for the diagnosis, however, radiographs, sialography, CT scan and MRI can help to confirm a diagnosis and localize a stone. commonest site of the calculi is the duct however it can be found inside the gland itself. non-invasive treatment techniques can be used such as shock-wave lithotripsy and endoscopic removal by sialendoscopy, surgery remains the technique of choice for intraglandular calculi.

**Conflict of interest:**

This research did not receive any specific grant from funding agencies in the public, commercial, or not-for-profit sectors.